\documentstyle{elsart}

\input epsf

\begin{document}

\begin{frontmatter}

\title{Improved energy resolution for VHE gamma-ray 
astronomy with systems of Cherenkov telescopes}

\author{W.~Hofmann$^*$},
\author{H. Lampeitl},
\author{A.~Konopelko},
\author{H.~Krawczynski}

\address{Max-Planck-Institut f\"ur Kernphysik, P.O. Box 103980,
        D-69029 Heidelberg, Germany}
\address{$^*$Corresponding author (Werner.Hofmann@mpi-hd.mpg.de\\
Fax +49 6221 516 603, Phone +49 6221 516 330)}

\begin{abstract}
We present analysis techniques to improve the energy resolution
of stereoscopic systems of imaging atmospheric Cherenkov
telescopes, using the HEGRA telescope system as an example.
The techniques include (i) the determination of the height
of the shower maximum, which is then taken into account in the
energy determination, and (ii) the determination of the location
of the shower core with the additional constraint that the
direction of the gamma rays is known a priori. This constraint
can be applied for gamma-ray point sources, and results in a
significant improvement in the localization of the shower core,
which translates into better energy resolution. Combining
both techniques, the HEGRA telescopes reach an energy resolution
between 9\% and 12\%, over the entire energy range from 1 TeV
to almost 100 TeV. Options for further improvements of the energy
resolution are discussed.
\end{abstract}

\end{frontmatter}

Imaging atmospheric Cherenkov telescopes (IACTs) represent
the prime instruments for gamma-ray astronomy in the TeV energy
range \cite{review}. With a number of sources established as TeV gamma-ray emitters
in IACT observations, emphasis is starting to shift from the pure
detection of sources to the precise determination of gamma-ray 
spectra. The energy of gamma rays is determined from the intensity of
IACT images, taking into account the radial distribution of
Cherenkov light within the light pool. In case of stereoscopic
systems of multiple IACTs, which observe an air shower from different
viewing angles, the location of the shower axis and hence the
distance of a given telescope from this axis
can be obtained by a simple geometrical
reconstruction. For single IACTs, the impact distance can be
estimated based on the location and shape of the Cherenkov image
within the camera, albeit with larger uncertainty. 
Energy resolutions quoted around 1 TeV 
for single telescopes vary between 29\%-36\%~\cite{whippleres,501whipple},
$\approx$~30-35\%~\cite{501hegra} and 20-28\%~\cite{catres}. The HEGRA
systems of IACTs provides a resolution of about 20\%~\cite{501hegrasys}.

Sources such as the Crab Nebula or the AGN Mkn 421 show spectra which
are consistent with pure power laws, $dN/dE \sim E^{-\alpha}$, with
spectral indices $\alpha$ ranging between 2.5 and about 4. In the
determination of power-law energy spectra, energy resolution is not
a very critical parameter. Convolution of a power-law spectrum with
a resolution function of constant width $\Delta E/E$ will result in
a spectrum with the identical spectral index. A correction is required
in the determination of the flux at or above a certain energy,
but this correction is modest even for instruments with a 
poor resolution $\Delta E/E \approx 40\%$.

The situation changes once sources exhibit a cutoff in the energy
spectrum, such as observed for Mkn 501 
\cite{501whipple,501cat,501array,501hegra,501hegrasys}. 
For the interpretation of the cutoff
phenomenon, e.g. in terms of absorption of gamma rays in 
interactions with the infrared/optical background, it is 
important to precisely map the shape of cutoff. Smearing of
the spectrum with an energy resolution in the 20\% range 
may distort its shape significantly. In principle, the original
spectrum can be recovered by unfolding techniques
(see, e.g., \cite{501hegra} and refs. given there). However,
all such techniques
result in rapidly increasing statistical
errors, once the bin size of the unfolded spectrum approaches
the energy resolution of the instrument -- after all, the loss
of information cannot be recovered and leads to this penalty.

It is therefore of significant importance to improve the
energy resolution of IACTs. In this article, we will demonstrate
that with new analysis techniques, a significant improvement
of the energy resolution of stereoscopic systems of IACTs can
be achieved, in particular if the source of gamma rays can be
considered a point source with known position. The results are
based on Monte-Carlo simulations of the HEGRA IACT system, but
they should apply in similar form to the various new systems
of IACTs which are currently planned or in construction.

\section{Factors governing the energy resolution of Cherenkov telescopes}

The energy resolution of Cherenkov telescopes is governed by a 
number of factors, among them
\begin{description}
\item[Statistical fluctuations in the image.] Since the number
of photoelectrons in a typical image is O(100), statistical
fluctuations in the number of photoelectrons limit the resolution
to O(10\%). Additional fluctuations arise from the amplification 
process in the photomultiplier and from night-sky background under
the image. In case of the HEGRA telescopes, the amplification noise
increases the fluctuations by a about a factor 1.2 compared to the Poisson
fluctuations alone, and the night-sky noise in a typical image
corresponds to about 4-5 p.e. rms.
\item[Image truncation.] In order to reduce the influence of
the night-sky background, the image intensity
is usually 
summed only over `image pixels' above a minimum intensity,
cutting away the tail of the image. 
The sum over image pixel amplitudes provides the so-called {\em size}
parameter used to derive the shower energy. Such a `tail cut' introduces
both additional noise as well as systematic nonlinearities;
for low-intensity images a larger fraction of the image is cut
than for intense images. An additional truncation occurs for
images which extend beyond the edge of the camera. At the 10\%-level,
edge effects start to matter at distances as large as 
$0.8^\circ$ between the image centroid and the edge of the camera.
\item[Threshold effects.] In the region near the trigger threshold
-- in case of the HEGRA telescopes this corresponds to images
with around 40 p.e. -- the image intensity detected in the camera
will be strongly biased, since showers with upward fluctuations
in the image {\em size} will have a larger probability of triggering.
In the sub-threshold energy region, the mean intensity of triggered
images will approach a constant, independent of the shower energy,
making an energy determination impossible.
\item[Errors in the localization of the shower core.] To convert
the measured image intensity into a shower energy, the distance
between the telescope and the shower axis needs to be known.
For the HEGRA system, the core is located with a precision of 10~m
to 20~m, depending on the core distance. In particular for
telescopes beyond the Cherenkov radius of about 120~m, where the
light intensity varies rapidly with core distance, the resulting
uncertainty in the energy estimate may exceed 30\%.
\item[Fluctuations in the shower development.] Variations in
the shower development provide a significant contribution to
the energy resolution; particularly relevant are fluctuations
in the height of the shower maximum, related primarily to the
fluctuation in the depth of the first interaction. Showers with
their maximum deeper in the atmosphere have a higher intensity
of light within their light pool, both because of the smaller
distance between the telescope and the light source, and 
because of the lower Cherenkov threshold at reduced height.
\item[Systematic errors.] All techniques for energy determination
rely heavily on Monte-Carlo simulations to provide the relation
between image parameters and shower energy, and to describe the 
performance of the telescope hardware. Imperfections in the
simulations of the air shower, or of the telescopes, or 
alignment errors and calibration errors not included in the
simulations may have a detrimental effect on the energy resolution.
Great care must be taken to ensure that the simulations properly
reproduce all relevant aspects of the data.
\item[Monte Carlo statistics.] Algorithms for energy reconstruction
frequently use multi-dimensional lookup tables to convert
values of image parameters into energy estimates. Given the 
time-consuming generation of Monte-Carlo events in particular
at the higher energies, the number of Monte-Carlo events is
frequently similar to, or even
inferior to the number of showers detected in the
experiment. Statistical errors in the table values may be
significant. They can be alleviated by an efficient choice of
variables, and by appropriate smoothing of the tables or fitting with
a smooth analytical function.
\end{description}

In this paper, we will concentrate on improvements of the energy 
resolution due to two factors, namely the experimental determination
of the height of the shower maximum on an event-by-event basis,
and an improved algorithm for the determination of the shower
core, applicable for gamma-ray point sources.

Additional improvements should be possible with an improved 
image analysis, e.g. by fitting image templates to the 
observed images to properly account for and compensate the
truncation effects mentioned above. The detailed discussion of such
effects goes beyond the scope of the current work.

\section{The HEGRA IACT system, its modeling, and data analysis}

The analysis techniques presented on the following have been
developed using Monte Carlo simulations of the HEGRA IACT system.
The HEGRA IACT system is located at 2.2 km asl. on the Canary
Island La Palma, at the Observatorio del Roque de los Muchachos.
The stereoscopic IACT system consists of five telescopes,
four arranged in the corners of a square with 100 m side length
and one in the center. Since the final telescope - one of the
corner telescopes -- was added rather late to the system, most
of the data taken so far use only four telescopes; the Monte Carlo
simulations were therefore also restricted to four telescopes. Each of the
HEGRA telescopes has 8.5~m$^2$ mirror area, 5~m focal length, and
is equipped with a 271~pixel camera with a 4.3$^\circ$ field of 
view. Shower signals in at least two telescopes are required
to trigger the system. Details about the telescopes, their trigger
system and their performance can be found in 
\cite{hermann_padua,perf_paper,trigger_paper,501hegrasys}.

The Monte Carlo simulations of the HEGRA telescopes are
described in \cite{hegra_mc}. The simulations provide a detailed
account both of the evolution of air showers and of the
telescope hardware, including a detailed modeling of the
optical path and the detection of photons, and of the
electronics signal processing by a 120 MHz Flash-ADC system.
Simulated events are passed through the same full reconstruction
chain as is used for real data.

In order to test the algorithms for the reconstruction of the
height of the shower maximum, the Monte Carlo was modified to
output the number of Cherenkov photons generated
as a function of atmospheric depth. 
To define the maximum emission, a smooth function was fit to
the depth profile. When the term `height of the shower maximum'
is used in the following, it refers to the height of maximum
emission of Cherenkov photons rather than to the height with 
maximum number of shower particles. Because of the variation of
atmospheric density with height, the maximum Cherenkov emission
occurs below the maximum in terms of particle number. Values
given for the shower height always refer to the height above
the telescopes.

Unless otherwise mentioned, studies were carried out for vertical showers.

The data analysis chain and the cuts are
similar, but not identical to those
used in \cite{501hegrasys}).
The location of the shower axis, required as input for any energy 
determination, was determined by geometrical reconstruction;
in case events are overconstrained (observation by three or more
telescopes), images were combined taking the errors on the image
parameters into account (Algorithm 2 of \cite{shower_reco}),
and cuts were applied on the resulting $\chi^2$ to reject the small
fraction of poorly reconstructed events. Only events
with cores reconstructed 
within 200~m from the central telescope were accepted.
For the energy 
reconstruction, only telescopes were used which fulfill the
following criteria: (i) at least 40 photoelectrons in the
image, (ii) the distance between telescope and shower core
does not exceed 200~m, and (iii) in order to exclude images
truncated by the edge of the cameras, the image centroid
had to be within $1.5^\circ$ from the camera center. At
least two telescopes had to be available for the energy 
determination.

\section{Influence of fluctuations in the shower height on the
energy determination}

\begin{figure}
\begin{center}
\mbox{
\epsfxsize10.0cm
\epsffile{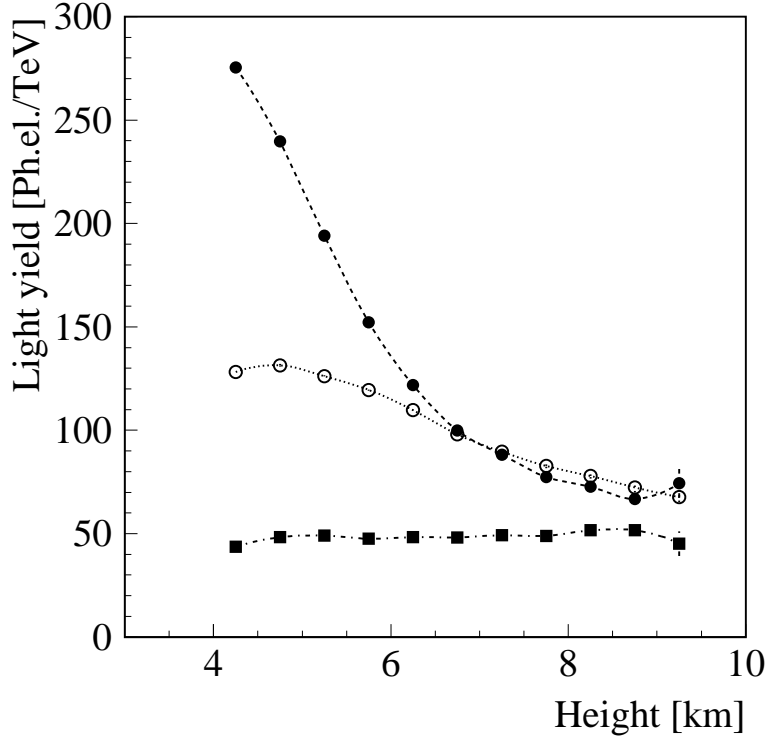}}
\end{center}
\caption{Light yield detected by the Cherenkov
camera as a function of the (measured) height of the shower maximum, a
distances of 40-50 m from the center of the light pool (full
circles), 90-100 m (open circles), and 140-150 m (full
squares), based on Monte Carlo simulations. The light yield
is given in units of photoelectrons/TeV.}
\label{fig_height}
\end{figure}

The influence of the height of the shower maximum 
on the Cherenkov light yield is illustrated 
in Fig.~\ref{fig_height}, which displays the light yield observed in
a Cherenkov telescope at different distances
from the shower axis as a function
of the height of the shower maximum above the telescopes. 
At TeV energies, the average height of the shower maximum is
about 6 km, with an
rms variation of 800~m, roughly corresponding to one
radiation length. The dominant contribution to this variation
comes from the fluctuation in the depth of the first interaction.
The influence of the
height of the maximum on the light yield is dramatic at small
distances from the shower axis, decreases smoothly out to the
Cherenkov radius of 120~m, and is modest at larger distances.

\begin{figure}
\begin{center}
\mbox{
\epsfxsize17.0cm
\epsffile{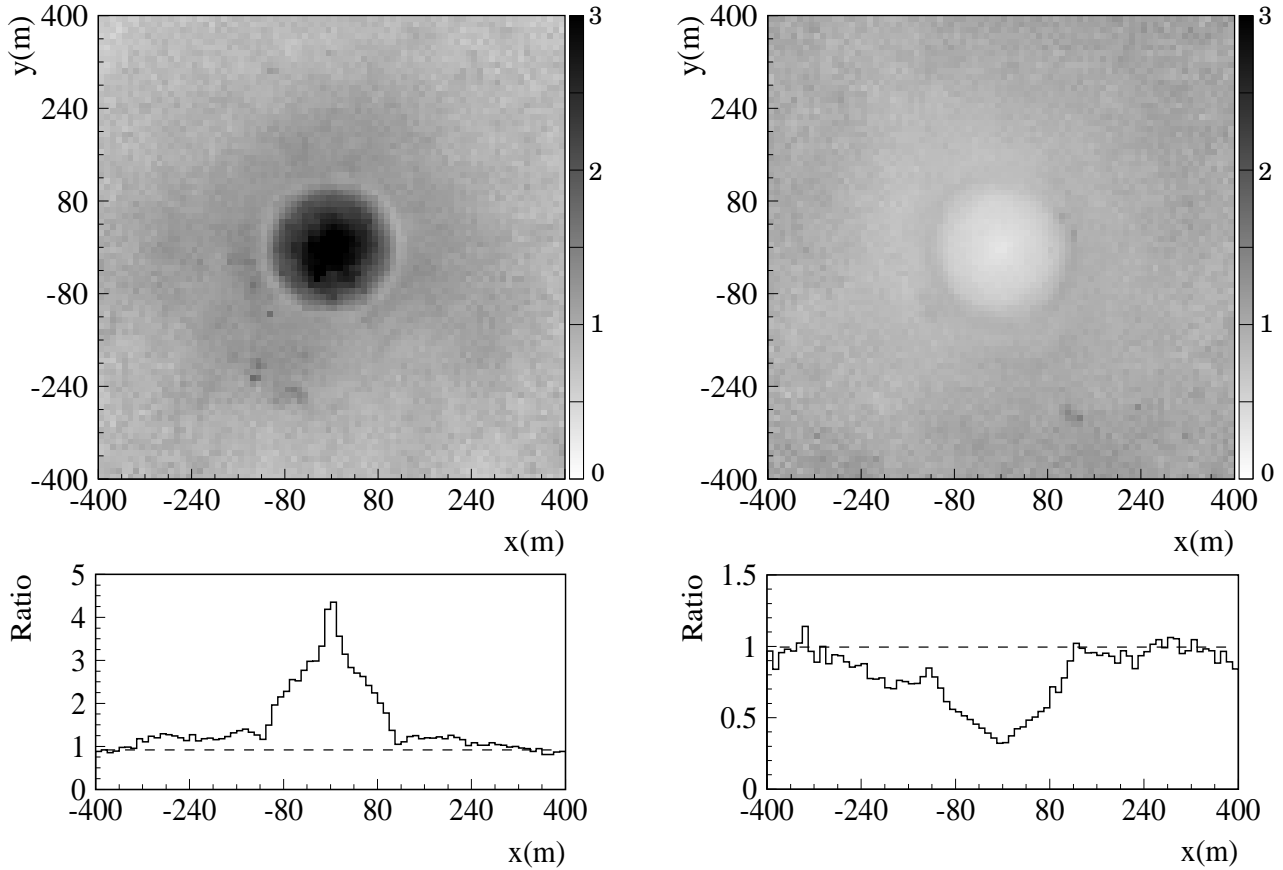}
}
\end{center}
\caption{Photon density on the ground for two selected 1 TeV
gamma ray showers, normalized to the average photon density.
The shower axis intersects the ground level at $x = y = 0$.
The lower panels show a cross section along the $x$ axis,
at $y = 0$.}
\label{fig_ex}
\end{figure}

The variation in the height of the shower maximum is by far the
dominant contribution to the fluctuations of the photon density
on the ground. This is illustrated in Fig.~\ref{fig_ex},
where the distribution of photons is shown for two simulated
showers, normalized to the mean photon density
obtained by averaging over many events. The first
event, with a low shower maximum, shows a strong enhancement
of photon density for radii below 80~m; in the second event,
with a high shower maximum, the photon density is reduced in this 
area. In both events, the distribution of photons is rather smooth
\footnote{Statistical fluctuations in the light yield are
negligible on this scale, given that each bin represents an
area of 10~m x 10~m and contains O($10^4$) photons per event.}
and symmetric about the shower axis, indicating that the 
height of the shower maximum influences the light distribution
in a very global fashion, and that other, more local fluctuations
are less important.
This observation implies that one should be able to rather
efficiently correct for the fluctuation of the shower
maximum, provided that this quantity can be measured on an
event-by-event basis.

In fact, the development of the analysis technique described
in this paper was triggered by the experimental observation of such 
global fluctuations of the light yield \cite{kruger_energy}:
for air showers observed simultaneously by all four of the
HEGRA telescopes, the energies determined independently by
two subsystems of two telescopes each agreed significantly
better than expected under the assumption of uncorrelated measurements,
indicative of a common factor influencing both measurements.

\section{Reconstruction of shower height}

To correct for the dependence of light yield on the height
of the shower maximum,
this height has to be determined from the information
contained in the multiple images. To a first approximation,
one may assume that the image of the shower maximum
coincides with the centroid of the Cherenkov image. Denoting
by $h_{max}$ the height of the shower maximum, by $r$ the
distance from the telescope to the shower axis, and by
$d$ the distance in the image plane between the centroid
of the image and the image of the source (in units
of degrees\footnote{Coordinates in the camera plane are expressed
in units of degrees and represent the slope of photon
trajectories relative to the telescope axis; they should not
be confused with the orientation angle of images within 
the camera.}), one finds
$$
{1 \over h_{max}} = {\pi \over 180^\circ}~{d \over r}~~~~~~~~.
$$
With a single image, $r$ and $h_{max}$ cannot be determined
separately. With a stereoscopic IACT system, $r$ is obtained
from the geometrical reconstruction of the shower axis 
and then an estimate for $h_{max}$ can be obtained for
each image, and be suitably averaged over telescopes. Note that
measurement errors tend to be Gaussian in $d$, hence one
should average the estimates of $1/h_{max}$.

\begin{figure}
\begin{center}
\mbox{
\epsfxsize8.0cm
\epsffile{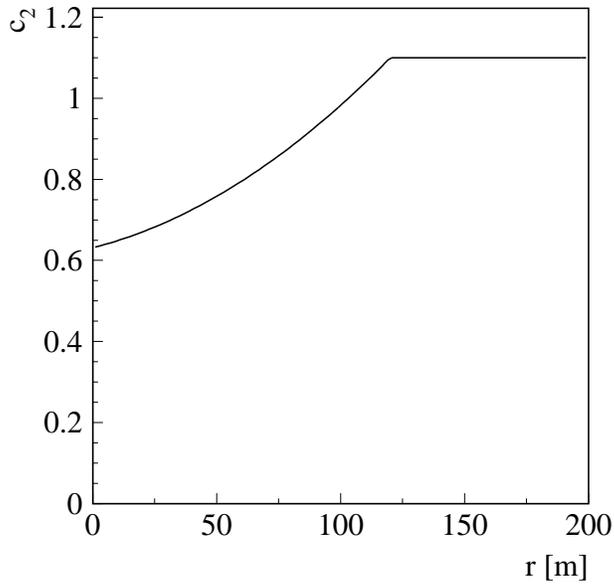}}
\end{center}
\caption{Coefficient $c_2(r)$ relating the height of
the shower maximum to the distance $d$ between the
image of the source and the image centroid.}
\label{fig_c2}
\end{figure}

Monte Carlo simulations show, however, that this model is
oversimplified; while light detected at large radii,
beyond the Cherenkov radius, is predominantly generated
around the shower maximum, light detected at smaller
distances from the shower core is generated deeper in the
atmosphere. In general, the relation between $d$ and
$h_{max}$ can be parameterized as
$$
{1  \over h_{max}} = c_1(r) + c_2(r) {\pi \over 180^\circ}~{d \over r}~~~~~~~~,
$$
with a small offset $c_1$ of at most 0.1/km
and $c_2$ rising at small $r$, leveling off
at about unity beyond the Cherenkov radius.
Tests showed that in the reconstruction of the 
shower height, $c_1$ could be neglected. For $c_2$,
a simple parameterization was used, see Fig.~\ref{fig_c2}.
The values determined for $1/h_{max}$ from the
individual telescopes were averaged, assigning
to $d$ an error
$$
\Delta d = \max \left( {2.5 \over \sqrt{I}}~,~0.15 \right)~[degr.]~~.
$$

\begin{figure}
\begin{center}
\mbox{
\epsfxsize10.0cm
\epsffile{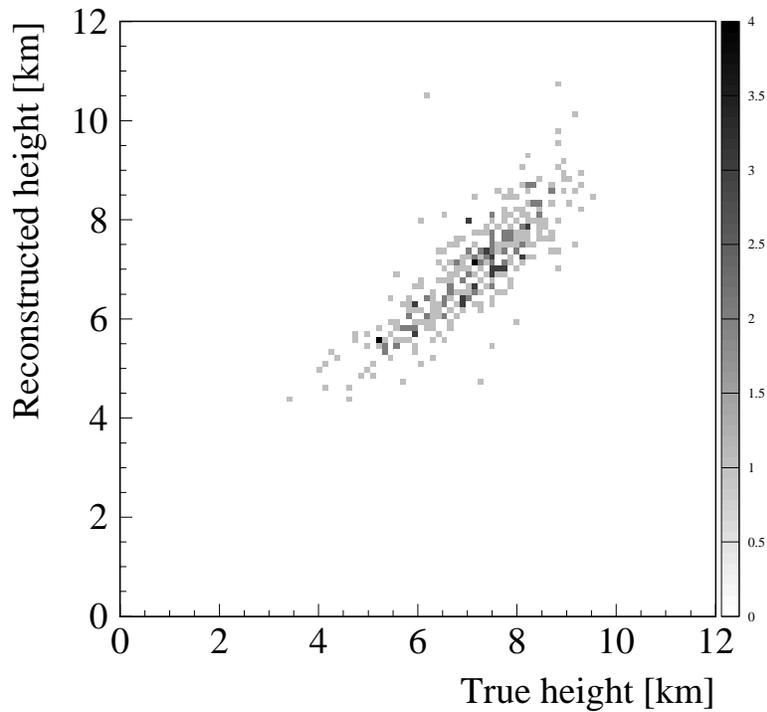}}
\end{center}
\caption{Distribution of the measured height of the shower
maximum vs. the true height, for simulated gamma-ray showers
at TeV energies.}
\label{fig_hres}
\end{figure}

Here, $I$ denotes the image {\em size} defined as the sum of the 
amplitudes of image pixels.
Fig.~\ref{fig_hres} illustrates the distribution of 
measured height of the shower maximum vs. the true height,
for events reconstructed
based on two or more telescopes. With the standard
reconstruction of the shower geometry, an rms resolution in the height of the
maximum of about 600~m is achieved, corresponding to
about 0.7 radiation lengths. With the improved determination
of the core location (see below), a resolution of about 530~m is obtained.

\section{Improved determination of core location}

\begin{figure}
\begin{center}
\mbox{
\epsfxsize10.0cm
\epsffile{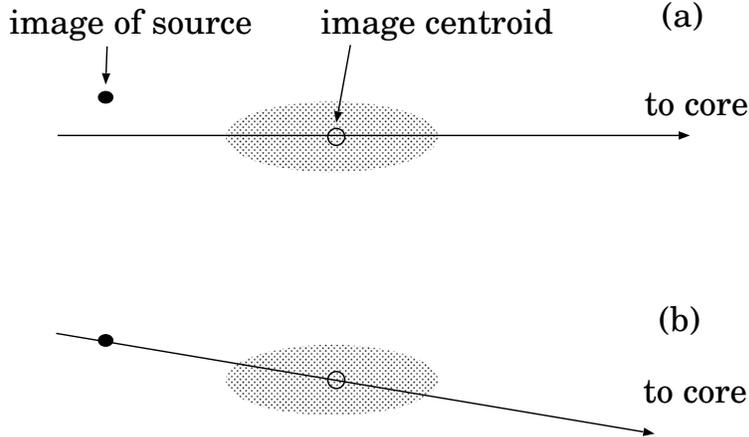}}
\end{center}
\caption{(a) Normal reconstruction of the core location
based on the assumption that the image axis points to the
location where the shower axis intersects the plane
of the telescope dish. With stereoscopic observation
of a shower by two telescopes, the impact point can
be determined by intersecting the image axes,
starting from the telescope locations. (b) Reconstruction of the core location
assuming that the gamma-ray comes from the known point 
source, and using the vector from the source to the 
centroid of the image to define the direction to the
core. In contrast to the image orientation, the image
centroid is usually quite well determined, and therefore
the technique (b) usually provides a much better estimate
of the direction to the core.
}
\label{fig_scheme}
\end{figure}

\begin{figure}
\begin{center}
\mbox{
\epsfxsize10.0cm
\epsffile{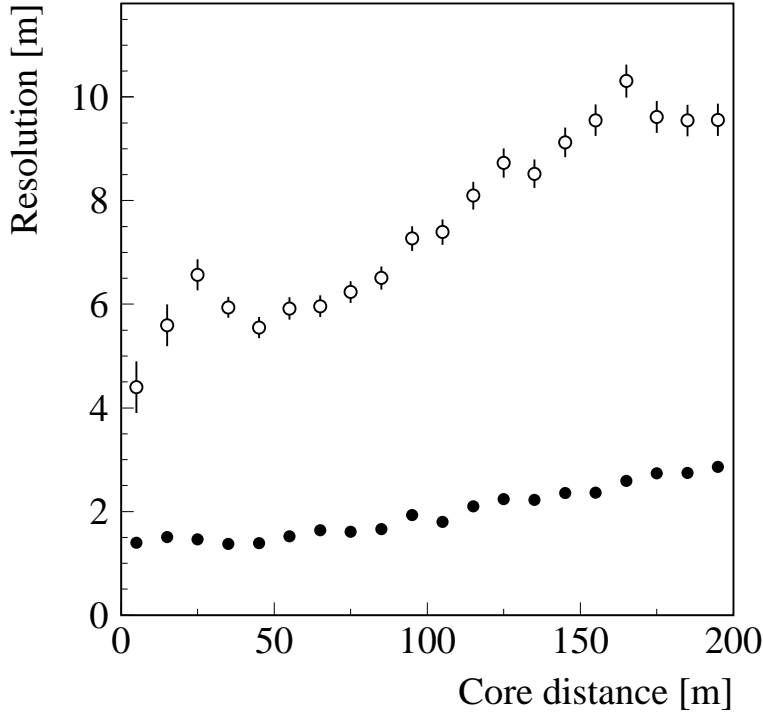}}
\end{center}
\caption{Resolution in the reconstruction of the
core location, projected onto one of the coordinate
axes, as a function of the distance of the shower
core from the central telescope of the HEGRA IACT
system. The simulations assume an energy spectrum
similar to the spectrum observed for Mkn 501. 
Open circles: reconstruction of both the shower
direction and the core location. Full circles:
reconstruction of the core location, assuming that
the shower direction is known.}
\label{fig_coreres}
\end{figure}

In the geometrical reconstruction of the geometry of air
showers on the basis of multiple stereo views, the 
dominating source of measurement errors is the
uncertainty in the determination of the orientation
of the images, with a typical error of about $4^\circ$ for gamma-ray
images containing 100 photoelectrons, detected in the
HEGRA cameras. The image centroid is located with a 
precision of $0.02^\circ$, in the direction transverse to the 
image axis. The high precision in the location of the
centroid opens the way to an improved determination of
the shower core, for gamma-rays emitted from a point source.
For a known source location, the image orientation 
can be recalculated using the measured position of the
image centroid, see Fig.~\ref{fig_scheme} 
(see also \cite{krennrich_padua}). With a typical distance of $1^\circ$
between the image of the source and the centroid of 
the Cherenkov image, the image orientation can be
derived with an uncertainty of about $1^\circ$, four times better
than from the image alone. These improved values
provide then the input for the determination of the
core location. The resolution in the core coordinates
is illustrated in Fig.~\ref{fig_coreres}; indeed, one finds
that the resolution, which normally varies between 6~m and 10~m in
each coordinate, is improved to values between 1.5~m and 3~m.

\section{Energy determination}

The standard procedure to reconstruct gamma-ray energies
using the HEGRA IACT system uses look-up tables to
relate the shower energy to the image {\em size} value $I$ measured at
a certain distance from the shower axis. The 
look-up tables store the mean {\em size} $I(E,r)$ for 18 bins in core distance,
and 16 bins in energy, derived from Monte
Carlo simulations. The relation between {\em size} and energy is
nonlinear, because of the effects of tail cuts and since the 
mean depth of the shower maximum varies with energy. For a given
event, for each telescope an energy value is derived by suitable
interpolation of the table values, and the average over telescopes
is formed. To account for the dependence on zenith angles,
the tables exist for four different zenith angles, and are interpolated.

\begin{figure}
\begin{center}
\mbox{
\epsfxsize11.0cm
\epsffile{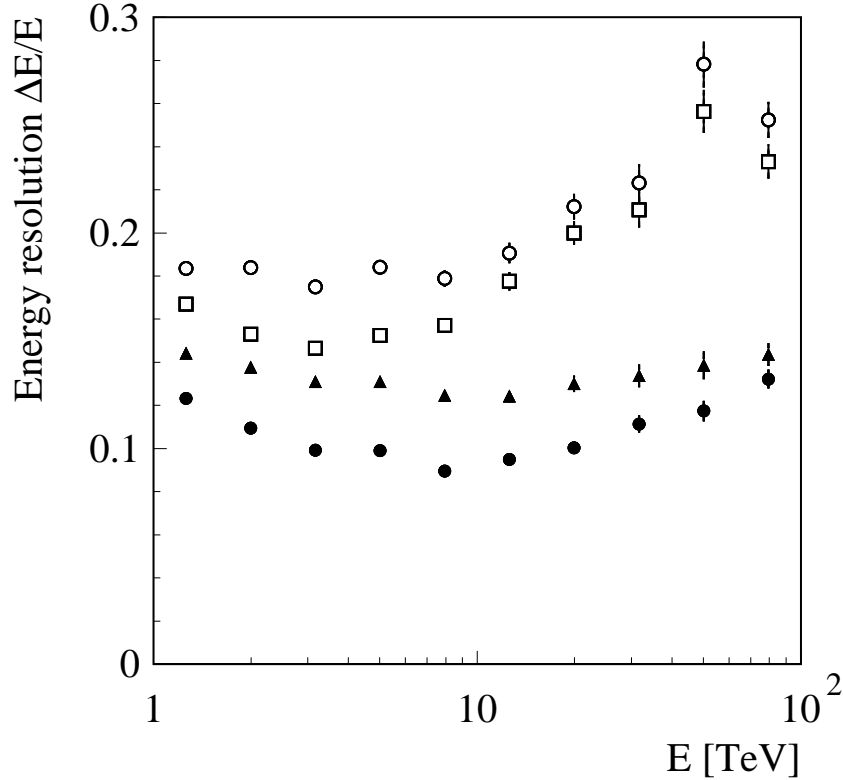}}
\end{center}
\caption{Energy resolution determined by fitting a Gaussian to
the ratio of reconstructed energy to true energy, as a function
of the energy of gamma rays. Open circles: conventional energy
determination. Full triangles: including the information on
the height of the shower maximum. Full circles: using in addition
in the core reconstruction the assumption that gamma-rays come
from a known point source. Open squares: conventional energy
determination, with the core reconstruction assuming a point 
source.}
\label{fig_eres}
\end{figure}

With this procedure, an energy resolution slightly below
20\% is achieved for energies up to 20~TeV (open circles in
Fig.~\ref{fig_eres}). At energies beyond 20 TeV, the resolution 
deteriorates slightly because of the increasing core distances.

Using the information on the shower height, one can change the
strategy completely: for a fixed height of the shower maximum,
the relation between energy and light yield at a fixed distance
from the core should be linear. Therefore, the energy 
corresponding to a given image {\em size} $I$ was determined as
$$
E = I~f(h_{max},r)~t(d_0)~g(I) ~~~.
$$
The basic (tabulated) function $f$ describes the {\em size}/energy
ratio as a function of (measured) shower height $h_{max}$ and radius $r$;
Fig.~\ref{fig_height} in fact shows the values of $1/f$ 
(in photoelectrons/TeV) for three
different ranges in $r$.
Additional correction functions $t(d_0)$ correct for image truncation
due to the camera edge, as a function of the {\em distance} $d_0$ from
the camera center, and for image truncation due to tail cuts, $g(I)$.
(Strictly speaking, the intensity lost due to tail cuts is a 
function not only of image {\em size} $I$, but also of the image 
shape, but for the current purpose the simple correction seemed
sufficient.) The energy resolution obtained with this procedure
(full triangles in Fig.~\ref{fig_eres}) is  improved significantly,
and varies between 12\% and 14\%.

Using the same procedure, but with the improved determination of the
core location on the basis of a known direction of gamma rays,
the energy resolution shown as full circles in Fig.~\ref{fig_eres}
is achieved, which now varies between 9\% and 12\%, almost a factor
two better than with the conventional energy reconstruction.
We note here that for background events not coming from the source,
a wrong energy value will be obtained. However, in a statistical
subtraction of background events using a suitable off-source data
sample, these events will be canceled no matter how the energy
estimate was obtained or how biased it is.

Of course, one can also combine the conventional energy reconstruction
with the improved core determination assuming a point source (open
squares in  Fig.~\ref{fig_eres}); while some improvement is observed,
it is clear that the correction for the varying shower height 
provides the bulk of the improvement, in particular at higher
energies.

The technique was also applied to showers at non-zero zenith
angles, using  events simulated at $20^\circ$, $30^\circ$ and
$45^\circ$ zenith angle. The lookup tables were generated separately
for each zenith angle, with an interpolation for intermediate values.
The reconstruction techniques - the combination of the shower-height
correction and the improved core loaction - work at all zenith
angles. At 10 TeV, e.g., which is well above threshold at all
 angles, the energy resolution is about $9\%$ independent of the zenith angle. 

\section{Concluding remarks}

We will first give some caveats, and then add some ideas for further
improvement of the energy resolution of IACT systems.

A key issue in the applicability of the energy 
determination assuming a point source of gamma rays
is the mode of failure in case the source
is extended, or has an extended component.
One will, of course, study the reconstructed 
angular distributions of gamma-rays; an extended
source should be recognized as such if its size
is comparable to the angular resolution of the telescopes,
of about $0.1^\circ$. If a somewhat extended source
is mistaken as a point source, the determination of 
the core location will suffer, and in the extreme case -
a source size equivalent to the angular resolution - the
gain in the core determination will be lost completely,
resulting in a core resolution equivalent to that provided
by the normal reconstruction. As a result, the technique provides
a soft failure mode in the sense that for marginally extended
sources the core resolution and hence the energy resolution
will degrade -- in the worst case back to the values without the
source constraint -- but the spectra will not be systematically
biased. 

Before applying the source constraint, one also needs to make
sure that without this constraint the source is reconstructed
exactly (within O($0.01^\circ$)) at the nominal location. Shifts
-- e.g. due to alignment problems -- could generate 
distortions of the spectrum. In general, the requirements
on the alignment of telescopes, both absolute and relative to
each other, are increased compared to the normal reconstruction.

Since the entire energy reconstruction is based on correction
functions derived from Monte Carlo simulations, one has to ensure
that these simulations are correct at the appropriate level of
precision. To achieve 10\% energy resolution in the actual data
sample, it is not sufficient to apply an algorithm which with
Monte Carlo events provides this resolution, but one must make 
sure that radial distributions etc. are indeed correctly
predicted, with systematic deviations well below the 10\%
target resolution. We believe that with the redundant information
provided by systems of three or more IACTs, and with the large gamma-ray
samples gained e.g. from Mkn 501, one will be able to verify the
simulations at such a level. Tests include the comparison of
gamma-ray parameters reconstructed using two subsystems of two
telescopes each \cite{kruger_energy}, 
and the comparison of the energies reconstructed
by two telescopes at different core distances, in analogy to the
technique used in \cite{pool_paper} to measure the radial
distribution of Cherenkov light. 

While the achieved 10\% energy resolution is certainly quite
acceptable, there are a number of ways further improvements
might be reached. For example, the corrections for image
truncation are certainly not optimal; using corrections which
are optimized for the 1 TeV range, rather than for the
entire sample of Monte Carlo events from a few 100 GeV to 
100 TeV, the resolution at 1 TeV can be improved by about 1-2\%.
This suggests an iterative approach, where a first energy
estimate is used to select correction functions optimized for
the corresponding energy range. Currently, the available
Monte-Carlo statistics limits the detail and dimension of
the correction tables. Alternatively, a fit to 
image templates \cite{hegra_fit,cat_fit} could be 
used to correct truncation. One could 
also try to avoid tail cuts entirely in the determination of
the image {\em size}, and generously add border pixels to the 
images.

The current technique, which uses the information
summarized in the Hillas image parameters, exploits 
for the energy
estimate only the
information on the height of the shower maximum. 
Using the full pixel information, one might try
to extract the full depth profile of the shower, and include
also the higher-order corrections.

In summary, the clear improvement in the energy resolution
of IACT systems demonstrates the wealth of information contained
in the multiple and redundant images of gamma-ray showers;
the techniques presented here should be seen as a first step 
towards an improved analysis, and almost certainly do not
represent the last word.

\section*{Acknowledgments}

The support of the HEGRA experiment by the German Ministry for Research 
and Technology BMBF and by the Spanish Research Council
CYCIT is acknowledged. We are grateful to the Instituto
de Astrofisica de Canarias for the use of the site and
for providing excellent working conditions. We thank the other
members of the HEGRA CT group, who participated in the construction,
installation, and operation of the telescopes, as well
as in the analysis and interpretation of  data.

\end{document}